\documentclass[11pt,a4paper]{article}
\usepackage{jheppub}
\usepackage{amsmath,amstext,amssymb}
\usepackage{graphicx}
\usepackage{feynmp}
\usepackage{epstopdf}
\usepackage{caption}
\usepackage{float}

\def\be{\begin{equation}}
\def\ee{\end{equation}}
\def\ba{\begin{eqnarray}}
\def\ea{\end{eqnarray}}

\newcommand{\bq}{ \textbf{q}}

\newcommand{\bk}{ \textbf{k}}

\usepackage{graphics}
\usepackage{graphicx}
\usepackage{dcolumn}
\usepackage{bm}
\usepackage{epsfig}
\usepackage{graphicx}
\usepackage{multirow}
\usepackage{dcolumn}
\usepackage{graphicx,epsfig}%

\begin{document}


\title{\center The Odderon in QCD with running coupling}

\author[a]{Jochen Bartels,}
\author[b]{Carlos Contreras,}
\author[c]{G. P. Vacca}

\affiliation[a]{II. Institut f\"{u}r Theoretische Physik, Universit\"{a}t Hamburg, Luruper Chaussee 149,\\
D-22761 Hamburg, Germany}
\affiliation[b]{Departamento de Fisica, Universidad Tecnica Federico Santa Maria, Avda.España 1680, Casilla 110-V, Valparaiso, Chile }
\affiliation[c]{INFN Sezione di Bologna, DIFA, via Irnerio 46, I-40126 Bologna, Italy}

\emailAdd{jochen.bartels@desy.de}
\emailAdd{carlos.contreras@usm.cl}
\emailAdd{vacca@bo.infn.it}

\abstract{Starting from the leading Odderon solution of the three gluon system in perturbative QCD we introduce, as a first step towards the transition to the nonperturbative region, an infrared cutoff and use the running QCD coupling constant. In our numerical analysis we find that the fixed cut solution with intercept one persists, hinting at a physical 
Odderon with intercept one and a  small t-slope. }

\today


\maketitle
 
\section{Introduction}

Recently TOTEM  \cite{Antchev:2017yns,Antchev:2017dia,Csorgo:2019fbf} data have stimulated \cite{oddetotem1}  a vivid discussion whether, in addition to the C-even Pomeron,  also a C-odd Odderon exchange is needed to describe the data.  After the proposal of Lukaszuk and Nicolescu ~\cite{Nicolescu} in 1973, it was the ISR data for $\frac{d \sigma}{dt}$  which indicated a difference between $pp$ and $p\bar{p}$ and hence raised the quest for a C-odd exchange at high energies. A first connection with QCD was made by Donnachie and Landshoff \cite{Donnachie:1983ff} who introduced a three-gluon exchange as a model for the Odderon.
   
In the early 80`s, soon after the discovery of the perturbative QCD Pomeron (BFKL)~\cite{BFKL}, which describes the composite state of two reggeized gluons it was realized that this picture can be generalized to composite states of three (and more) reggeized gluons, the so-called BKP states  ~\cite{Bartels:1980pe,Kwiecinski:1980wb}. 
A first solution of the three gluon problem was found by Janik and Wosiek \cite{Janik:1998xj} and its intercept  was found to be $ \alpha_O= 1 - 0.24717\frac{\alpha_s N_c}{\pi}$, which for a realistic  $ \alpha_s= 0.2$ yields  $ \alpha_O= 0.96$. In 1999 another solution was found by Bartels, Lipatov, and Vacca  ~\cite{Bartels:1999yt} with 
intercept exactly at one, $ \alpha_O= 1$, independent of the value of $\alpha_s$. A remarkabkle feature of this solution of the three gluon composite state equation is that it coincides with the two gluon BFKL solution with conformal spin  $n=1$. A discussion of the relevance of the JW and  the  BLV solutions in phenomenology prior to LHC data can be found in  \cite{Bartels:2001hw, ewert1}.

These perturbative results cannot directly be applied to soft hadron-hadron scattering. However, in recent years some progress has been made in analyzing the transition from the perturbative BFKL Pomeron to the soft Pomeron.
Starting from the perturbative region and replacing the fixed coupling by the running coupling, first an infrared cutoff has to be introduced.  These steps lead to important changes of the energy spectrum: for fixed coupling the BFKL  Pomeron has a fixed (i.e. $t$-independent) cut in the $\omega$ plane (angular momentum $j=\omega+1$), starting at 
$\omega_{cut}=  \frac{N_c \alpha_s}{\pi}\, 4 \ln 2$ 
and extending to $-\infty$.  In the presence of an infrared cutoff and with running $\alpha_s$  the piece of the $\omega$-cut between $\omega_{cut}$ and zero is replaced by an infinite sequence of discrete poles, which accumulate at zero. 
This picture has been verified in numerical studies,  for several different versions of an infrared cutoff: in \cite{Braun:1996tc} an infrared cutoff has been introduced in such a 
way that the BFKL bootstrap property (related to $s$-channel unitarity) is preserved;
in \cite{Kowalski:2017umu,Kowalski:2015paa,Kowalski:2014iqa}) boundary values of the BFKL amplitude are imposed at a fixed momentum scale $k_0^2$; in~\cite{Levin:2014bwa, Levin:2015noa, Levin:2016enb} a Higgs mass is introduced as an IR regulator, and in  \cite{Bartels:2018pin} a more sophisticated regulator is introduced which allows to embed the BFKL Pomeron into RG flow
equations. Details of this discrete spectrum in the $\omega$-plane of course depend upon the value of the cutoff scale and vary from one scheme to another, but the qualitative picture is the same in all schemes. 
Next, for this discrete part of the spectrum also the eigenfunctions  have been studied  \cite{Bartels:2018pin}: most important, it has been found that only for the leading eigenvalue the wave function is centered in the 'soft' region of small transverse momenta, whereas for the nonleading
eigenvalues the wave functions become 'hard', i.e. these Pomeron states are centered in the UV-region of large transverse momenta. Consequently, their couplings to hadron states are expected to be small.
Finally, the $t$-slopes  \cite{Bartels:2018pin} of these discrete poles are largest for the leading eigenvalue, and go to zero for the nonleading poles. These findings suggest that these two steps - introduction of an infrared cutoff and of  the running coupling - bring us substantially closer to the nonperturbative region, in particular the existence of a 'soft'  Pomeron state with intercept above one. What remains is the 'unitarization' of this set of Pomeron states: this requires, in particular, the introduction of the triple Pomeron vertex. Work along this line is in progress. 

Applying these findings for the BFKL Pomeron now to three gluon problem of the Odderon, it seems plausible to 
proceed in the same manner: introduce an infrared cutoff and the running coupling and then study the energy spectrum. 
As already stated before, the leading BLV Odderon solution without IR cutoff and with fixed coupling leads to a fixed
(i.e. $t$-independent) cut in the $\omega$ plane, starting at $\omega=0$  and extending to $- \infty$. 
In this paper we will investigate how this picture changes, once we introduce an IR cutoff and the running coupling. For simplicity we use the Higgs-mass regulator, and we use the numerical methods outlined in \cite{Bartels:2018pin}. As the main result, we find that the spectrum remains unchanged, i.e. we still have a cut starting at $\omega=0$. The wave functions are
'hard', i.e they have their main support in the region of large transverse momenta, and the $t$-slopes are small.  
An analysis of what happens to the other family of Odderon solutions (JW) with lower intercept is unfortunately much more involved,  and it is extremely difficult to carry on employing a similar approach.

The  paper will be organized as follows. In section 2 we review the BFKL kernel with Higgs mass regulator for $n=1$ in the forward direction, and after introducing  for the fixed coupling case the lattice approximation  we present numerical results for the eigenvalue spectrum and for the eigenfunctions. This part is mainly meant to verify that our lattice approximation is consistent with our knowledge of the analytic BLV solution.
In section 3 we turn to the running coupling case and compute eigenvalues, eigenfunctions and $t$-slopes. In a final section  we summarize and discuss our results.

\section{The $n=1$ BFKL kernel with a Higgs mass regulator}

In this section we present the BFKL kernel with an infrared cutoff. This problem has been addressed before in  previous papers
\cite{Bartels:2018pin} and \cite{Levin:2014bwa,Levin:2015noa,Levin:2016enb}, and partly we follow those papers. 

In our   previous  paper \cite{Bartels:2018pin} we perform a numerical study  of the BFKL kernel for the Pomeron case 
with two  infrared regulator. In our analysis we  consider  both  the  Wilsonian optimized IR regulator in the  exact functional renormalization group approach (this regulator was constructed in such a way that the BFKL Pomeron becomes part of the exact renormalization group equations in the Multi Regge Kinematics)  and then we carried out 
a numerical study of the BFKL Pomeron with a "gluon mass" regulator.
In both cases we  computed the energy eigenvalues (i.e. poles in the
angular momentum plane), in particular intercepts and $q^2$ slopes of the Regge trajectory functions
and eigenfunctions of the BFKL kernel. From our results for the Wilsonian regulator and the mass regulator qualitatively there are no difference and 
then the general behavior  is independent of the regulator.

Then in this   sections we shall perform a numerical analysis of the IR modified BFKL
kernel   introducing a simple mass regulator.
First we will present  the  
BFKL kernel for the    fixed QCD coupling, and in a second step we also
consider a running gauge coupling. Our main focus is on the spectrum of the integral
kernel: eigenvalues, eigenfunctions, and $\bq^2$ slopes of the Odderon case. In the Pomeron case, we found a set  of  discrete
spectrum so that one can  make a link at large distances with the local Pomeron fields of a Reggeon Field Theory (RFT). 
Therefore we shall look for evidence of such a case for the Odderon.
We remind that the properties of both Pomeron and Odderon as a RFT, including their universal properties,
have been recently investigated using functional renormalization group methods in~\cite{Bartels:2015gou,Bartels:2016ecw}.
The numerical analysis proceeds in two steps. First we  study the eigenvalues and  eigenfunctions of the BFKL
Odderon  equation with the   mass regulator, then what  is new in our analysis are the $\bq^2$ slopes of
the odderon  states.  In a future paper   we turn to the Wilsonian IR regulator and, again,
compute those relevant properties.

\subsection{The $n=1$ BFKL equation in the forward direction}

We begin with the Higgs mass regulated BFKL kernel with fixed coupling. 
First we define:
\be
\label{mom-param}
\bq_1=\frac{\bq}{2} +\bk,\,\, \bq_2=\frac{\bq}{2}-\bk,\,\,\bq'_1=\frac{\bq}{2}+\bk,'\,\, \bq'_2=\frac{\bq}{2}-\bk'\,.
\ee
The analytic expression  of the symmetrized BFKL  kernels (the real part gluon emission)
 has the form
 \ba
\frac{2\pi}{ \bar{\alpha}_s} K(\bq,\bk,\bk')= \sqrt{\frac{\bq_1^2+m^2}{\bq_2^2+m^2}} \frac{1}{(\bk-\bk')^2+m^2} \sqrt{\frac{\bq_2'{}^2+m^2}{\bq_1'{}^2+m^2}}\nonumber\\
+\sqrt{\frac{\bq_2^2+m^2}{\bq_1^2+m^2}} \frac{1}{(\bk-\bk')^2+m^2} \sqrt{\frac{\bq_1'{}^2+m^2}{\bq_2'{}^2+m^2}}\nonumber\\
-\frac{\bq^2+\frac{N_c^2+1}{N_c^2} m^2}{\sqrt{(\bq_1^2+m^2)(\bq_2^2+m^2)(\bq_1'{}^2+m^2)(\bq_2'{}^2+m^2)}}
\ea
where  $\bar{\alpha}_s=\frac{N_c \alpha_s}{\pi}$, 
and the  gluon trajectory function (virtual part of the BFKL kernel) has the form:
\ba
\omega_g(\bk^2)=-\frac{\bar{\alpha}_s}{4\pi} \int d^2 k' \frac{\bk^2+m^2}{({\bk'}^2+m^2)((\bk-\bk')^2+m^2)}\nonumber\\
= -\frac{\bar{\alpha}_s}{2\pi}\int d^2 k' \frac{\bk^2+m^2}{({\bk'}^2+m^2)({\bk'}^2+(\bk-\bk')^2+2m^2)}\,.
\ea
The full BFKL kernel is then given by:
\be
\tilde{K}(\bq,\bk,\bk')= K(\bq,\bk,\bk')+\delta^{(2)}(\bk-\bk')  \left( \omega_g(\bq_1^2) +\omega_g(\bq_2^2) \right)\, .
\ee 

We first consider the forward direction $\bq^2=0$ where the kernel simplifies:
\ba
\label{Higgs-forwardkernel}
\frac{2\pi}{ \bar{\alpha}_s} K^{(0)}(\bk,\bk')= \frac{2}{(\bk-\bk')^2+m^2} - \frac{\frac{N_c^2+1}{N_c^2} m^2}{(\bk^2+m^2)({\bk'}^2+m^2) }
\ea 
and 
\be
\tilde{K}^{(0)}(\bk,\bk')= K^{(0)}(\bk,\bk')+2 \delta^{(2)}(\bk-\bk') \omega_g(\bk^2)\,.
\ee
The eigenvalue equation takes the form:
\be
\label{BFKLK}
 \omega f_{\omega}(\bk)= \frac{\bar{\alpha}_s}{2\pi}   \int d^2 \bk' \tilde{K}^{(0)} (\bk,\bk')f_{\omega}(\bk') 
\ee
In this paper we are interested in eigenvalues and eigenfunctions with conformal spin $1$:
\be 
\label{n=1eigenfunctions}
f(\bk) = e^{i\varphi} \tilde{f}(|\bk|)
\ee
where $\varphi$ is the azimutal  angle of the vector $\bk$.
Leaving the  forward direction  and including the $\bq^2$ dependence of the eigenvalues $\omega(\bq^2)$ we decompose into intercept $ \omega(0)$ and  $\alpha' $  the  slope:
\be
\label{trajexp}
\omega (\bq^2)  = \omega (0) +\alpha' \bq^2.
\ee

For our numerical analysis of the eigenvalue equation it will be convenient to combine terms which contain 
the potentially singular denominator $1/(\bk-\bk')^2$ and to rewrite the eigenvalue equation
in the following form\cite{Bartels:2018pin}:
\ba
\label{BFKL}
\omega f(\bk) & =&\frac{\bar{\alpha}_s}{2\pi} \int d^2k' \Big[ 
\frac{2 f(\bk')({\bk'}^2+m^2)- 2f(\bk) (\bk^2+m^2)}{({\bk'}^2+m^2)((\bk-\bk')^2+m^2)} - \frac{\frac{N_c^2+1}{N_c^2} m^2}{(\bk^2+m^2)({\bk'}^2+m^2) } f(\bk')\Big] \nonumber\\
 &&+\frac{\bar{\alpha}_s}{2\pi} \int d^2k'\frac{2 f(\bk) (\bk^2+m^2)}{({\bk'}^2+m^2)({\bk'}^2+(\bk-\bk')^2+2m^2)}.
\ea
This form has an integrand behaving manifestly better at large momenta $|\bk|\sim |\bk'|\to \infty$.
We  are interested in eigenfunctions of the form (\ref{n=1eigenfunctions}) and consider the following form of the eigenvalue equation:
\ba
\omega \tilde{f}(|\bk|) & =&\frac{\bar{\alpha}_s}{2\pi} \int d^2k' \Big[ 
\frac{2 \tilde{f}(|\bk'|)e^{i(\varphi'-\varphi)}({\bk'}^2+m^2)- 2\tilde{f}(|\bk|)| (\bk^2+m^2)}{({\bk'}^2+m^2)((\bk-\bk')^2+m^2)}\Big] \nonumber\\
 &&+\frac{\bar{\alpha}_s}{2\pi} \int d^2k'\frac{2 \tilde{f}(|\bk|) (\bk^2+m^2)}{({\bk'}^2+m^2)({\bk'}^2+(\bk-\bk')^2+2m^2)}
 \Big].
\ea
Here $\varphi$ and $\varphi'$ denote the azimutal angles of the vectors $\bk$ and $\bk'$, resp.
The angular integrations can be done by using the formulae
\be
\frac{1}{2\pi} \int_0^{2\pi} d \varphi \frac{1}{a+b \cos \varphi} = \frac{1}{\sqrt{a^2-b^2}}.
\ee 
and
\ba
\label{ang-int}
&&\frac{1}{2\pi} \int_0^{2\pi} d \varphi \frac{e^{i\varphi}}{a+b \cos \varphi}
= \frac{1}{2\pi} \int_0^{2\pi} d \varphi\frac{\cos \varphi}{a+b \cos \varphi}
=\frac{-b}{a+\sqrt{a^2-b^2}}    \frac{1}{\sqrt{a^2-b^2}}\,,
\ea
where 
\be
a=k^2+{k'}^2+m^2,\,\, b=-2 k k'.
\ee
Introducing the  short hand notations
\be
D=\bk^2 +m^2,\, D'={\bk'}^2+m^2,\,D''={\bk''}^2+m^2 
\ee
and
\ba
&&S_0=\sqrt{(k^2-{k'}^2)^2+2m^2(k^2+{k'}^2)+m^4},\nonumber\\
&&S_1=k^2+{k'}^2+m^2+S_0\nonumber\\
&& S_2=\sqrt{(k^2-{k'}^2)^2+2({k'}^2+2m^2) (k^2+{k'}^2)+({k'}^2+2m^2)^2}
\ea
the eigenvalue equation can be written as:
\ba
\label{n=1scalar}
\omega f(k) & =&\bar{\alpha}_s \int_0^{\infty} d {k'}^2  \Big[ \frac{2kk'}{S_1} \frac{D'}{D' S_0} f(k') -\frac{D}{D'S_0}f(k)\Big] \nonumber\\
&&+ \bar{\alpha}_s \int_0^{\infty} d {k'}^2 \frac{D}{D'S_0}
\ea

\subsection{Numerical results for eigenvalues and eigenfunctions for fixed coupling}

The numerical analysis of the eigenvalue equation (\ref{n=1scalar}) is done in the same way as described in \cite{Bartels:2018pin}:
 for the integration over ${k'}^2$ we introduce a lattice. First 
we change to logarithmic variables $t'=\ln \frac{{\bk'}^2}{m^2}$ with  $d{\bk'}^2 =dt' {\bk'}^2$ and then introduce a lattice in the new variables $t'$. Introducing the limits $k^2_{min}=10^{-40},
t_{min}=\ln \frac{k^2_{min}}{m^2}$ and $k^2_{max}=10^{80}, t_{max}= \ln \frac{k^2_{max}}{m^2}$
and dividing the interval $\Big[t_{min}, t_{max}\Big]$
into $N_{step}=600$ equal steps, we define the lattice points 
\be
t_i=t_{min}+i \frac{t_{max}-t_{min}}{N_{step}},\,\, k_i^2=m^2 e^{t_i}, \,\,i=0,...,N_{step}
\ee
and arrive at the discrete vector $f_i=f(k_i)$ and the discrete matrix $K_{ij}=K(k_i,k_j)$.
For the diagonal element we encouter the combination:
\be
\frac{-a+\sqrt{a^2-b^2}}{b}-1 = -4 k k' \frac{(k-k')^2+m^2}{(k-k')^2+m^2 +\sqrt{(k^2+{k'}^2+m^2)^2- 4 k^2 {k'}^2}}
\ee

In the following  we present our numerical results of the odderon eigenvalues and the wave functions. In this section we stick to the fixed coupling $\alpha_s$. 
It is convenient to introduce
\be
E_n=-\omega_n\,.
\label{energies}
\ee
We find discrete positive eigenvalues $E_n$, the largest one being very close to zero. The first three values are:
\be
E_1=0.000032,  \,\,\,\, E_2=0.000289,   \,\,\,\, E_3=0.000802
\ee
In Fig. 1 we present the first 30 eigenvalues of the Odderon with fixed coupling constant:
\begin{figure}[H]
\begin{center}
\epsfig{file=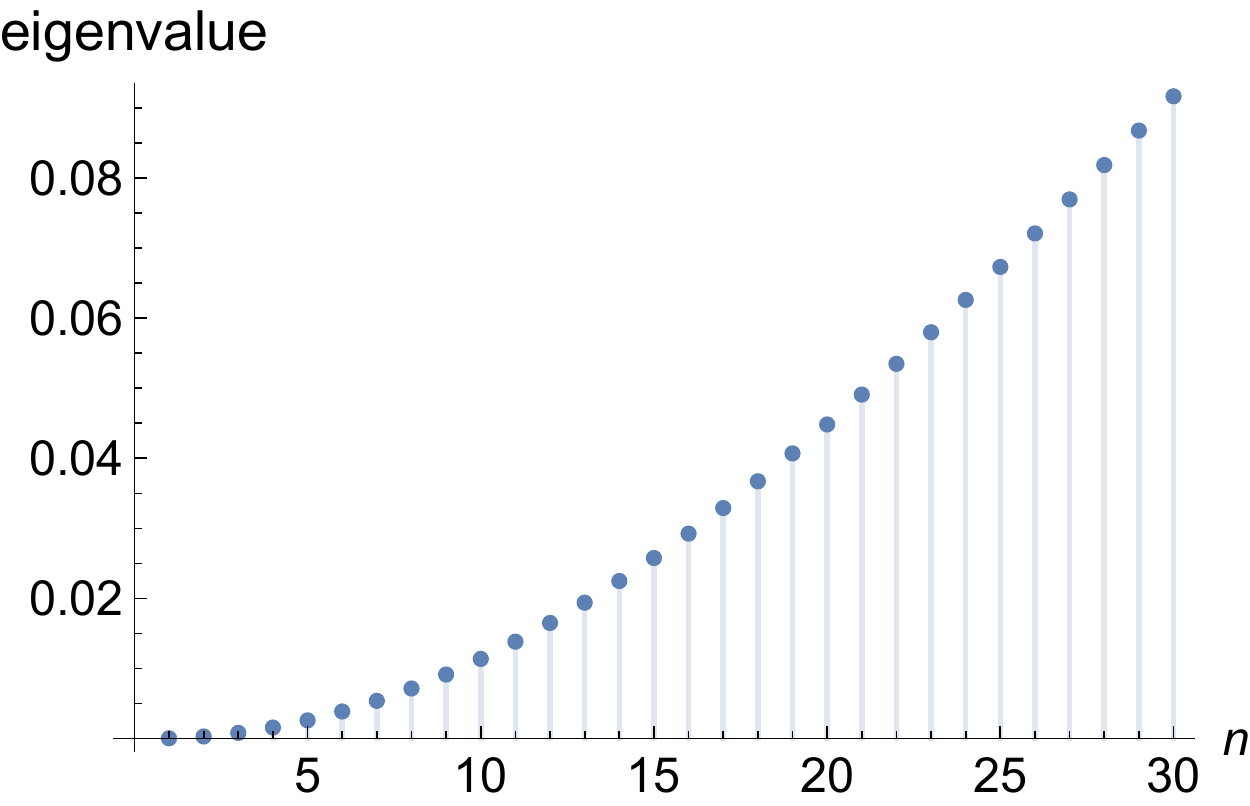,width=12cm,height=5cm}\\
\caption{The first 30 eigenvalues  of the Odderon with fixed coupling\label{Fig1}} 
\label{Fig1}
\end{center}
\end{figure}
We interpret these eigenvalues as being the lattice approximation of a cut in the positive energy plane, starting at zero.
As to the eigenfunctions, we find that they oscillate: the leading one has one maximum, the second one has 
one zero and has two extrema  etc. The oscillations extend over the full extension of the lattice provided that $ q^2> m^2$. For example, for the leading eigenvalue, the single node  has its center (on the logarithmic scale)  approximately  at  $57$, i.e. far in the UV region. In Fig.2 we show the first three eigenfunctions: 
\begin{figure}[H]
\begin{center}
\epsfig{file=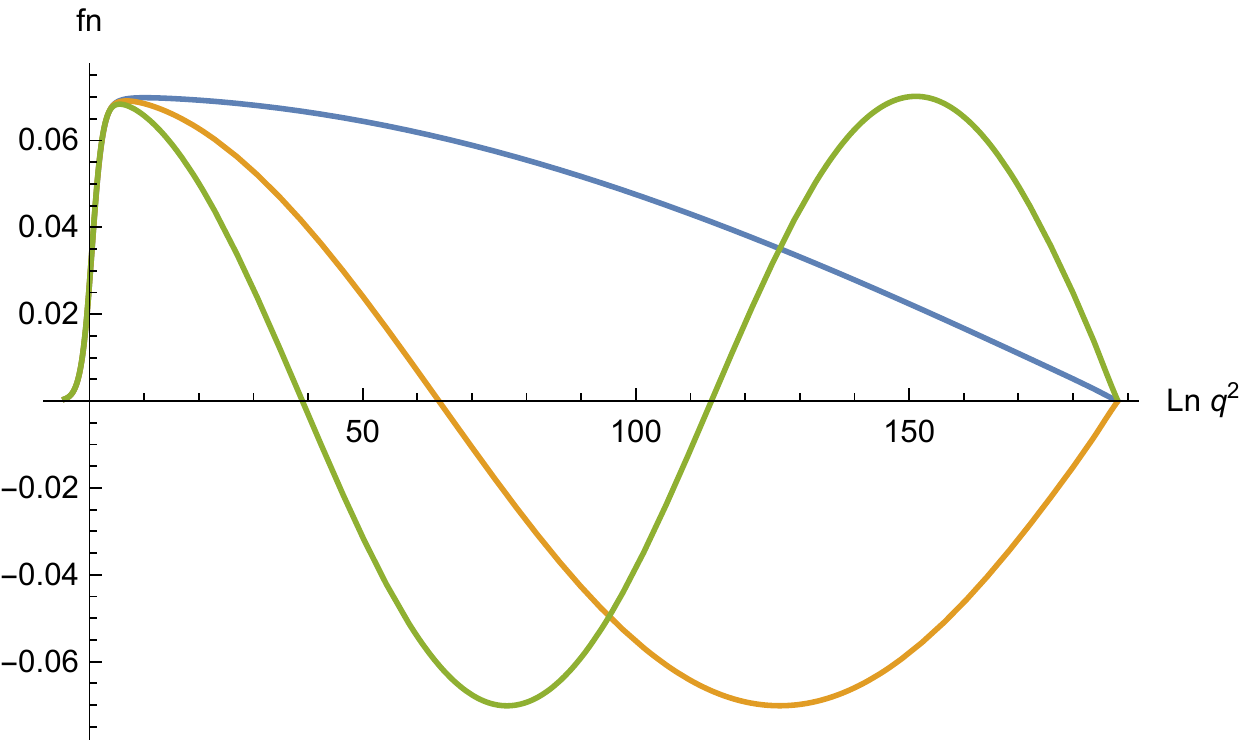,width=12cm,height=5cm}\\
\caption{The first three wavefunctions for fixed  coupling, as a function of $\ln q^2$ 
\label{Fig2}} 
\end{center}
\end{figure}

To make the support of the wavefunctions a bit more quantitative, we define the logarithmic radius    
\be
 < \ln \bq^2> = \frac{\int d\bk^2 |f_n(\bk)|^2 \ln \bk^2}{\int d\bk^2 |f_n(\bk)|^2}\,,
\ee
where momenta are in units of $m=0.54$ GeV. By exponentiating this logarithmic radius we translate these logarithmic radii to the linear scale (in units of GeV). For the lowest eigenvalues we find for the logarithmic radii
\ba
\label{rlog-1}
<ln \, k^2> _1&=&   56.75  \nonumber\\
<ln \, k^2> _2&=&90.39  \nonumber\\
<ln \, k^2> _3&=& 93.04 \, ,
\ea
and for the linear radii
\ba
\label{rlin-1}
r_1 &=&  1.14 \times 10^{12}\, GeV  \nonumber\\
r_2 &=&  2.30 \times 10^{19}\, GeV  \nonumber\\
r_3 &=&  8.62 \times 10^{19}\, GeV .
\ea
More general, in Fig.3 we show, for the first 20 eigenfunctions,  the logarithmic  and linear radii:
\begin{figure}[!h]
	\centering
	\begin{minipage}[t]{6cm}
		\centering
		\includegraphics[scale=0.5]{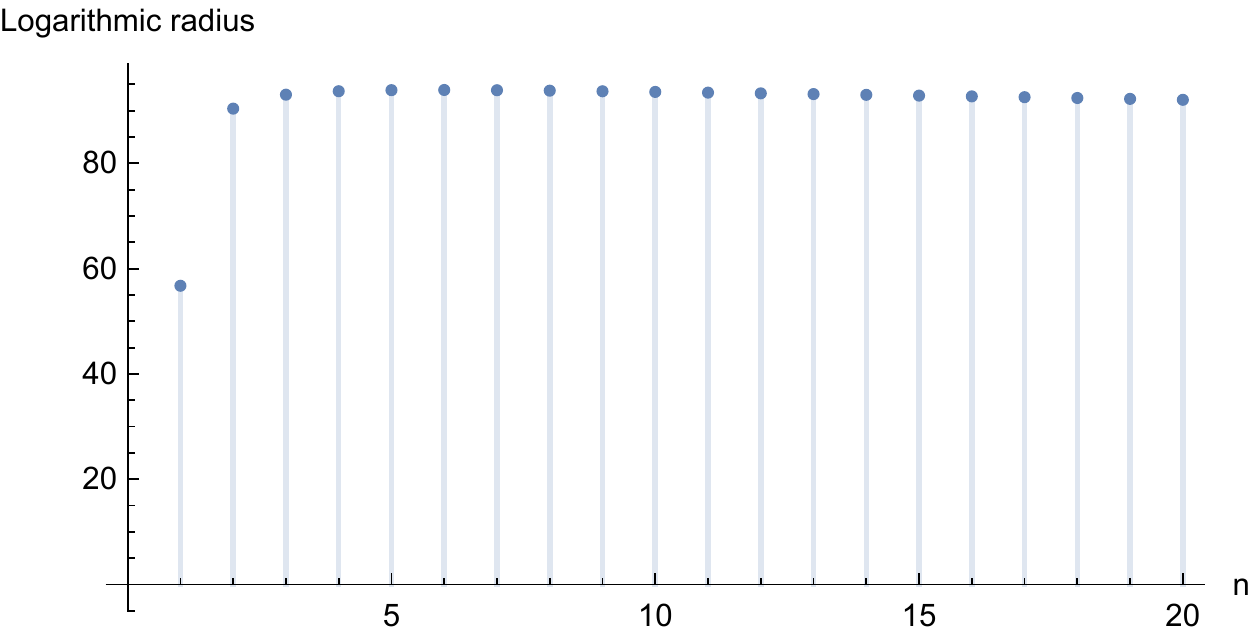}
	\end{minipage}
	\hspace{2cm}
	\begin{minipage}[t]{6cm} 
		\centering
		\includegraphics[scale=0.5]{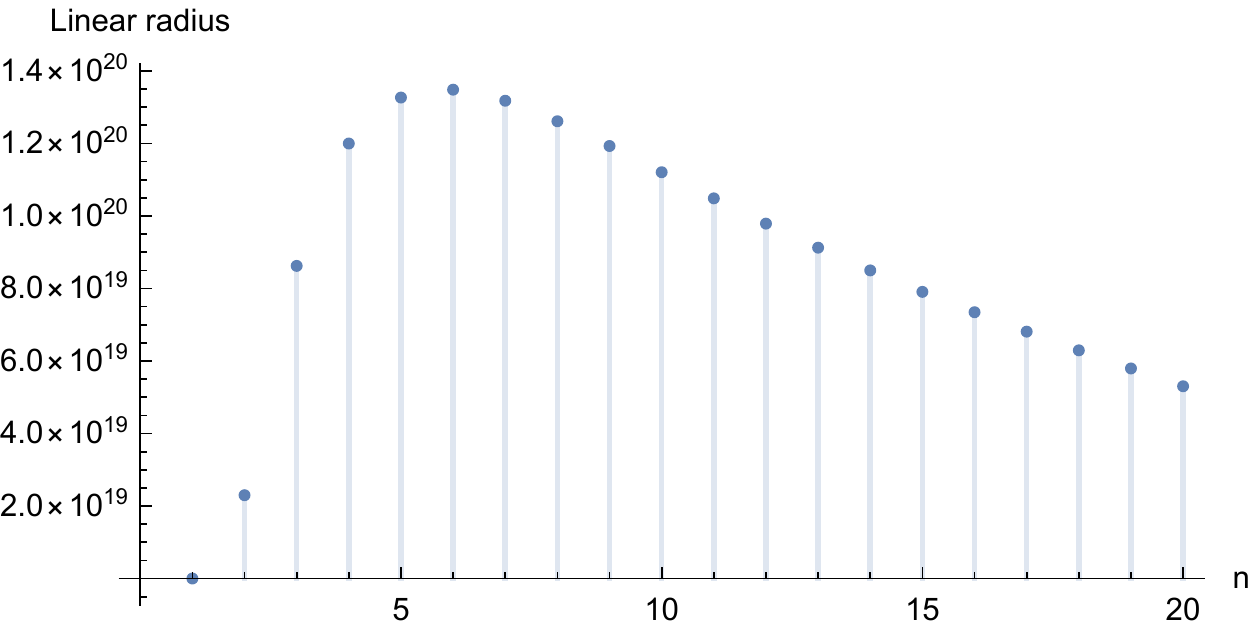}
	\end{minipage}
\caption{logarithmic (left) and linear (right) radii for the first 20 eigenfunctions}
\label{Fig3}
\end{figure}

\section{The Odderon solutions for the running coupling constant}
\subsection{Introducing the running coupling and leaving the forward direction} 

Let  us now turn to the  case of the running coupling. We follow the discussion of our previous paper \cite{Bartels:2018pin}. As a first step we 
simply replace the fixed coupling $\alpha_s$ by
\be 
\label{runningalpha}
\alpha_s(\bq^2) = \frac{3.41}{\beta_0 \ln (\bq^2+R^2_0)}
\ee
and
\be
\bar{\alpha}_s(\bq^2) = \alpha_s(\bq^2) \frac{N_c}{\pi}
\ee
with $\beta_0=(11 N_c - 2 N_f)/12$, $N_f=3$. Its normalization is chosen to match the measured value at the $Z$ mass scale.
$R_0$ defines the scale below which the running coupling is 'frozen'. Both $q^2$ and $R^2_0$ are in units of $\Lambda^2_{QCD}$, and $R_0$  has to be well above  
$ \Lambda_{QCD}^2=0.15^2$ GeV$^2$. In our calculations we use $R_0=0.54$ GeV. More accurate models allowing for different number of flavors  can be easily considered.
In our numerical computations with the Higgs regulator we actually find it convenient to 
follow the conventions used in \cite{Levin:2014bwa,Levin:2016enb}: we 
define momenta and $R_0$ in units of the regulator mass $m=m_h=0.54 GeV$.
This leads to the modification of (\ref{runningalpha}):
\be 
\label{runningalpha-pract}
\alpha_s(\bq^2) = \frac{3.41}{\beta_0 \Big[ \ln (\bq^2+R^2_0)+\ln \frac{m_h^2}{\Lambda^2_{QCD}}\Big]}
\ee
with $R_0=1$.  With this convention in all our previous expressions the mass $m=m_h$ will be replaced by unity.  

The inclusion of the QCD running coupling effects in the Regge limit is a delicate issue when considering a full resummation.
Strictly speaking this effect goes beyond the Leading Log contribution in the MRK, since one has to take into account emissions of at least two real gluons close in rapidity,
which start from the region called quasi multi regge kinematics. It is also well known that the BFKL Pomeron in NLL accuracy  has a spectrum which must be cured in the collinear regions with subleading term, and several approaches have been proposed. 
The same situation can be observed for the QCD perturbative Odderon, for which the kernel is also known to the NLL accuracy~\cite{Bartels:2012sw} 
and a solution with intercept at one is also expected~\cite{Bartels:2013yga}, at least in the large $N_c$ limit.

There is, however,  a consensus that a good understanding of the pure running coupling effects can be nevertheless obtained by directly improving 
the picture obtained from the leading logarithmic approximation, that is by simply replacing the fixed coupling by  a running coupling, 
even if this approach is not unique~\footnote{Another possible approach preserving the bootstrap property as in~\cite{Braun:1996tc} is considered elsewhere~\cite{MB and GPV2019}.}.
We shall take this attitude and consider in our calculation, the following prescription:\\
(i) in the trajectory function $\omega_g(\bq^2)$ we simply put
\be
\alpha_s \to \alpha(\bq^2)\,.
\label{traj-running}
\ee
\\
(ii) in real kernel $K_{\text{BFKL}}(\bq,\bq')$ in the forward direction is modified by the substitution
\be
\label{forward-kernel-running}
\alpha_s \to \sqrt{\alpha(\bq^2)\alpha({\bq'}^2)}.
\ee
\\
(iii) In the nonforward direction the kernel $K_{\text{BFKL}}(\bq_1,\bq_2;\bq'_1,\bq'_2)$ will be multiplied by
\be
\label{kernel-running}
\alpha_s \to \left(\alpha(\bq_1^2)\alpha(\bq_2^2)\alpha(\bq'_1{}^2)\alpha(\bq'_2{}^2)\right)^{1/4}.
\ee

As discussed before, we will  consider this prescription  as a first approximate attempt to include the running coupling  and for the forward direction the eigenvalue equations will be modified in the following way:
\be
\label{coupling-q2}
 \alpha_s({\bk'}^2) K(\bk',\bk'') \rightarrow  \sqrt{\alpha_s({\bk'}^2)} K(\bk',\bk'')\sqrt{ \alpha_s({\bk''}^2)},
\ee
and the trajectory functions will be simply multiplied by  $\alpha_s({\bk}^2)$.

Finally, for the $t$-slopes we have to leave the forward direction. In addition to the $\bq^2$ expansions of the kernel and of the trajectory function described in section 7.2 of \cite{Bartels:2018pin}, we also need the expansion of the running couplings in (\ref{traj-running}) and (\ref{kernel-running}). 
This situation is a bit more complicated, and both in the expansion (\ref{coupling-q2})
and (\ref{trajexp}) terms linear in $\bq$ have to be kept.  
However, as pointed out, the slope is relatively small for fixed coupling constant and one expects  that the running correction are more  relevant  for  the eigenvalues and eigenfunctions  but not for the slope.
In this approximation, we are now ready to present numerical results for the eigenvalues and for the slopes.

Next let us take a closer look at the dependence of the kernel on the momentum transfer $\bq^2$.
Again we start from \cite{Bartels:2018pin}, section 7.2.
The $\bq^2$ slopes of the eigenvalues are obtained from
\be
\label{slopes}
\omega_n(\bq^2)=\omega_n^{(0)}+\bq^2 \frac{\int d^2 \bk \int d^2 \bk' f_n({\bk'})\Big[ K^{(1)}(\bk,\bk')+2\delta^{(2)}(\bk-\bk')\omega_g^{(1)}(\bk^2)\Big]  f_n(\bk)}{\int d^2\bk |f_n(\bk)|^2}\,,
\ee
where $\omega_n^{(0)}$ are the eigenvalues of the forward kernel $K^{(0)}$, $f_n(\bk)$ the corresponding Odderon eigenfunctions, and $K^{(1)}$, $\omega^{(1)}$ the corrections
of the order $\bq^2$ to the forward BFKL kernel and the gluon trajectory, resp. 

In order to find $K^{(1)}(\bk,\bk')$ we expand the kernel in the small $\bq^2$ region to first order in $\bq^2$:
\be
\label{kernelexp}
K(\bq,\bk,\bk')=K^{(0)}(\bk,\bk')+\bq^2 K^{(1)}(\bk,\bk').
\ee
With the shorthand notations
\ba
D=\bk^2+m^2,\,\, D'={\bk'}^2+m^2,\,\,D_0=(\bk-\bk')^2+m^2
\ea
we find:
\ba 
\label{kernelexp}
K (\bq,\bk,\bk')= \frac{ \bar{\alpha}_s}{2\pi}\Big[  \frac{2}{D_0} \left(1-\frac{(2\bq\bk)(2\bq \bk')}{4D D'}+\frac{(2\bq\bk)^2}{8D^2}+\frac{(2\bq\bk')^2}{8{D'}^2}\right)\nonumber\\
-\frac{m^2 \frac{N_c^2+1}{N_c^2}}{DD'}  
\left(1+\frac{1}{2} (\frac{ \bq \bk}{D})^2 +\frac{1}{2} (\frac{ \bq \bk'}{D'})^2
-\frac{\bq^2}{4}(\frac{1}{D}+\frac{1}{D'})\right)-\bq^2 \frac{1}{DD'}\Big]\,.
\ea 
Note that there are no terms of the order $\bq$. 

For the integration over the azimuthal angles in (\ref{slopes}) we have to observe the angular dependence of the wave functions which leads to the additional factor
\be
e^{i(\varphi-\varphi')},
\ee
where $\varphi$ and $\varphi'$ denote the angle of the vectors $\bk$ and $\bk'$, resp.   We immediately see that 
for the terms in the second line of (\ref{kernelexp}) the angular integrations give zero. In the first line 
we use 
\ba
&&\frac{1}{(2 \pi)^2} \int_0^{2\pi} d\varphi  \int_0^{2\pi} d\varphi' e^{i(\varphi-\varphi')}
\frac{(2\bq \bk) (2\bq \bk')}{(\bk-\bk')^2 + m^2}\nonumber\\
&&=2 q^2 k k' \frac{k^2+{k'}^2+m^2}{S_0 S_1}
\ea
and
\ba
&&\frac{1}{(2 \pi)^2} \int_0^{2\pi} d\varphi  \int_0^{2\pi} d\varphi' e^{i(\varphi-\varphi')},
\frac{(2\bq \bk)^2 }{(\bk-\bk')^2 + m^2}\nonumber\\
&&=2 q^2 k^2 \frac{2kk'}{S_0 S_1}
\ea

With these expressions we find:
\ba
\label{K^1}
&&\frac{1}{2 \pi} \int_0^{2\pi} d\varphi  \int_0^{2\pi} d\varphi' e^{i(\varphi-\varphi' )}  K^{(1)} (\bk,\bk) \nonumber\\
&&=\bar{\alpha}_s \frac{2kk'}{S_1}  \frac{1}{S_0D D'}  \left( -\frac{m^2}{2} \right)\Big[
1+\frac{(k^2-{k'}^2)^2}{DD'}\Big]
\ea

For the $\bq^2$- expansion of the trajectory function we have the same expressions as for the Pomeron case, since
the delta functions $\delta^{(2)}(\bk-\bk')$ lead to $e^{i(\varphi-\varphi)} \to 1$. For our final result we use 
eq.(7.45) of  \cite{Bartels:2018pin} with $K^{(1)}$ from (\ref{K^1}).

\subsection{Numerical results}

We begin with the eigenvalues. Again we introduce the energies $E_n=-\omega_n$ and find a sequence of positive eigenvalues starting at 
\be
E_1= 4 \times 10^{-6},    \,\,\,\, E_2 = 28 \times 10^{-6}, \,\,\,\, E_3 = 74 \times 10^{-6}
\ee
which we interpret as approximating a cut in the positive energy plane starting at zero. The first eigenvalues are shown in Fig.4 :
\begin{figure}[H]
\begin{center}
\epsfig{file=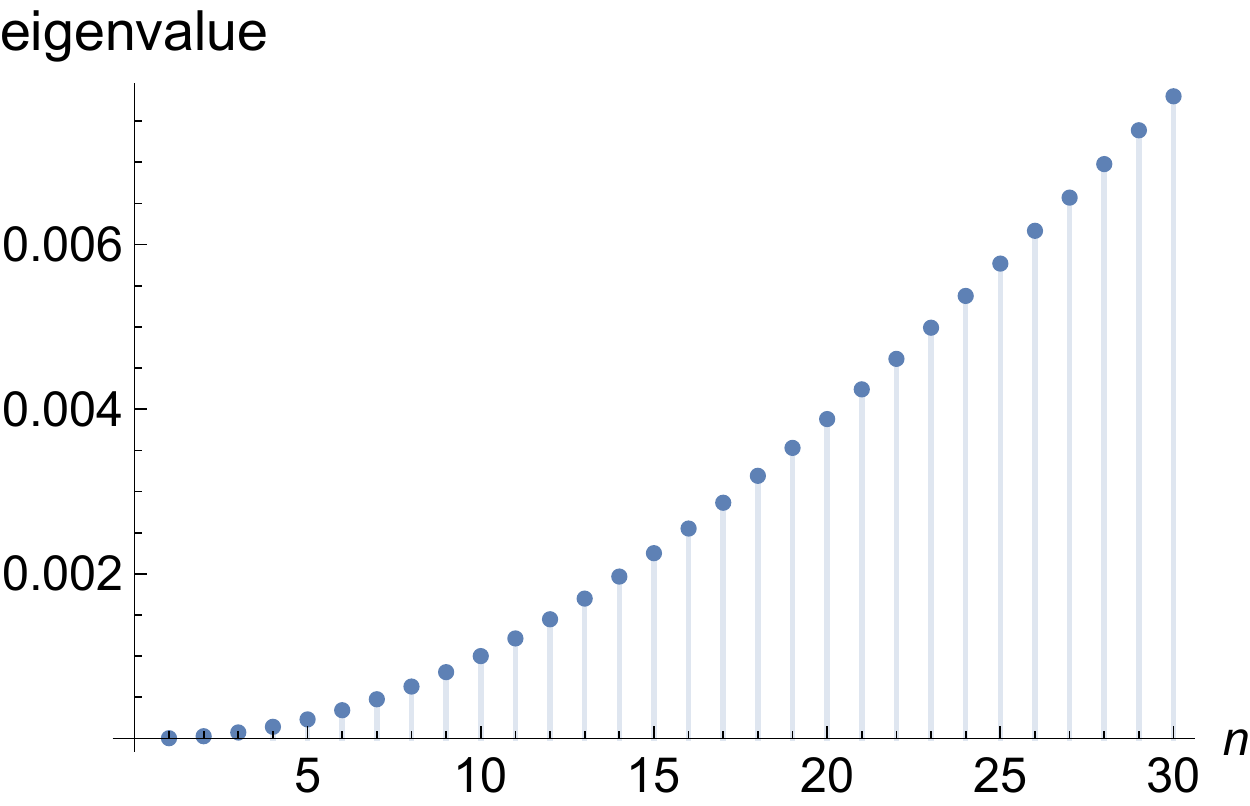,width=12cm,height=5cm}\\
\caption{The first 30 eigenvalues
\label{Fig4}} 
\end{center}
\end{figure}
\noindent
The curve in Fig.4 keeps the shape of Fig.1 and is only shifted a little bit. 

For the eigenfunctions we find that they again oscillate with same behaviour, extending over the full lattice region $ q^2> m^2$:
\begin{figure}[H]
\begin{center}
\epsfig{file=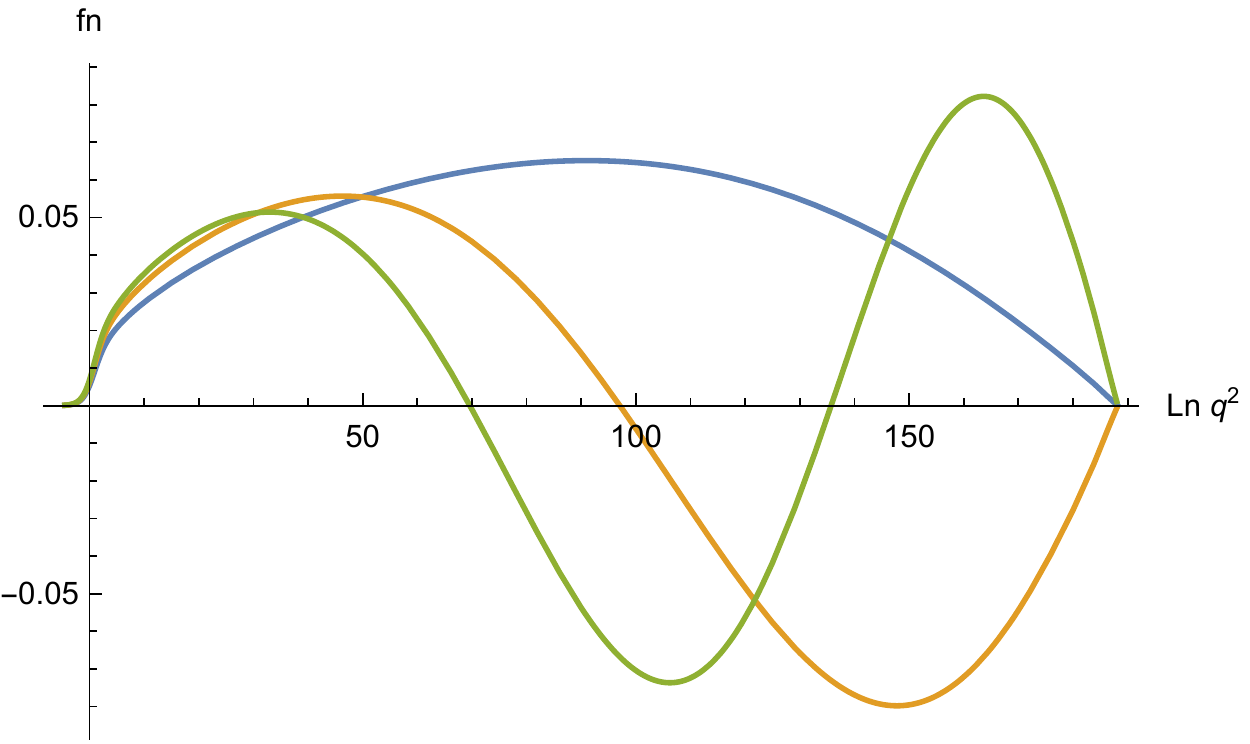,width=12cm,height=5cm}\\
\caption{The first three wavefunctions as a function of $\ln q^2$ for the runnuing coupling constant
\label{Fig5}} 
\end{center}
\end{figure}
\noindent
The effect of the running coupling constant is mainly  to smoothen the behaviour of the wave function, as one can see from  the Fig. 2 and Fig. 5  and to shift the center of them to the right. For example, for the leading eigenvalue, the single node  has its center (in the logarithmic scale)  approximately  $88$, compared with $57$ for the fixed coupling case:
\ba
\label{rlog-1}
<ln \, k^2> _1&=&   87.97  \nonumber\\
<ln \, k^2> _2&=& 107.26  \nonumber\\
<ln \, k^2> _3&=& 109.86 \, ,
\ea
which translates into the linear radii
\ba
\label{rlin-1}
r_1 &=&  6.83 \times 10^{18}\, GeV  \nonumber\\
r_2 &=&  1.06 \times 10^{23}\, GeV  \nonumber\\
r_3 &=&  3.89 \times 10^{23}\, GeV .
\ea
More general:
\begin{figure}[!h]
	\centering
	\begin{minipage}[t]{6cm}
		\centering
		\includegraphics[scale=0.5]{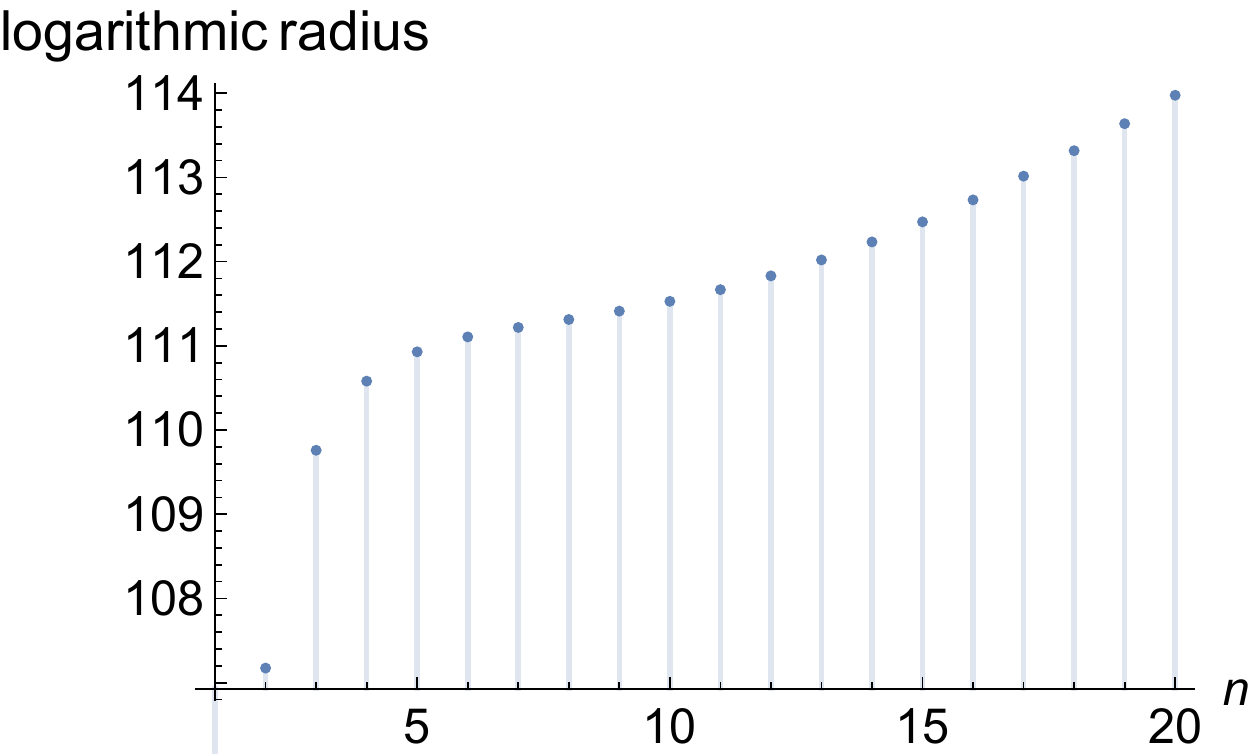}
	\end{minipage}
	\hspace{2cm}
	\begin{minipage}[t]{6cm} 
		\centering
		\includegraphics[scale=0.5]{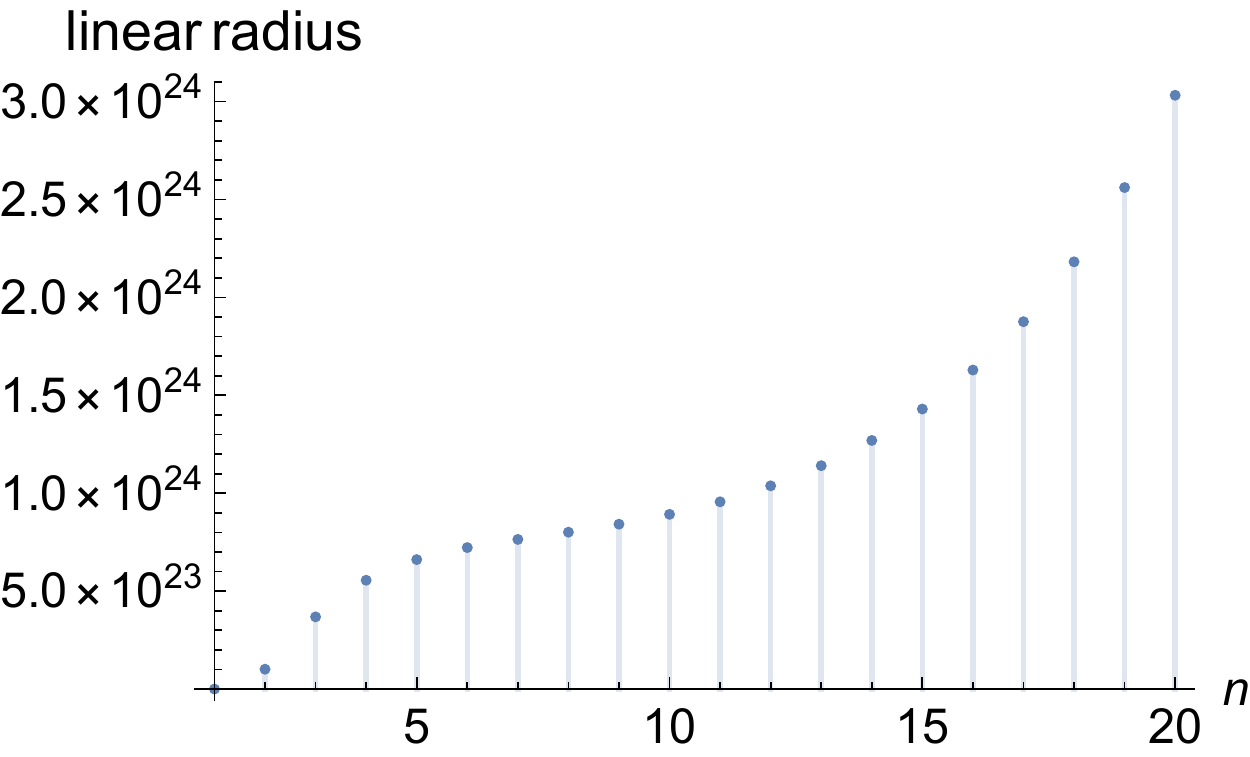}
	\end{minipage}
\caption{logarithmic (left) and linear (right) radii for the first 20 eigenfunctions for the running coupling}
\label{Fig6}
\end{figure}

It maybe useful to remember that for the massless case the BFKL eigenfunctions in the forward direction (for the symmetrized BFKL kernel) are given by
\be
f(\bk) \sim (k^2)^{-1/2-i \nu} e^{i n \varphi}
\ee
Near the beginning of the cut at $\omega=0$ we have $\nu=0$. Our lattice eigenfunctions have to be compared with $\sqrt{k^2}f(k^2)$: 
our leading eigenfunctions should therefore be seen as the lattice approximation of 
\be
 \sqrt{k^2}f(\bk) \sim (k^2)^{-i\nu} \,\,e^{i\varphi}.
 \ee
\noindent 
Since we are introducing a mass  as regulator of the infrared region, we expect that the wave function  is suppressed in the region  $k^2 < m^2$, and the form 
$ (k^2)^{-i\nu}$ is valid only for larger values of $k^2$. Putting  $t= \ln k^2$, we find that the wave function can be described approximately by:
\be
f_n(\bk) \sim \cos   \, \nu_n ( t-t^*) \hspace{0.5cm}  \text{or } \hspace{0.5cm} f_n(\bk) \sim \sin  \, \nu_n ( t-t^*) \,\, \, \text{for}  \, \, t>t^*(m).
 \ee
For the first and  second eigenfunctions we find that it is well described by the $ \sin   \, \nu_1 ( t-t^*)$  and $ \sin   \, \nu_2 ( t-t^*)$  with $\nu_1=\frac{\pi}{2(t_{max}-t^*)} = 0.008$ and  $\nu_2=\frac{\pi}{t_{max}-t^*} = 0.016$. In Figure 7 one can see the behavior for leading wavefunction obtained with our mass regulator compared with the oscillatory behaviour of the massless case of the BFKL functions.

\begin{figure}[!h]
	\centering
	\begin{minipage}[t]{3cm}
		\centering
		\includegraphics[scale=0.5]{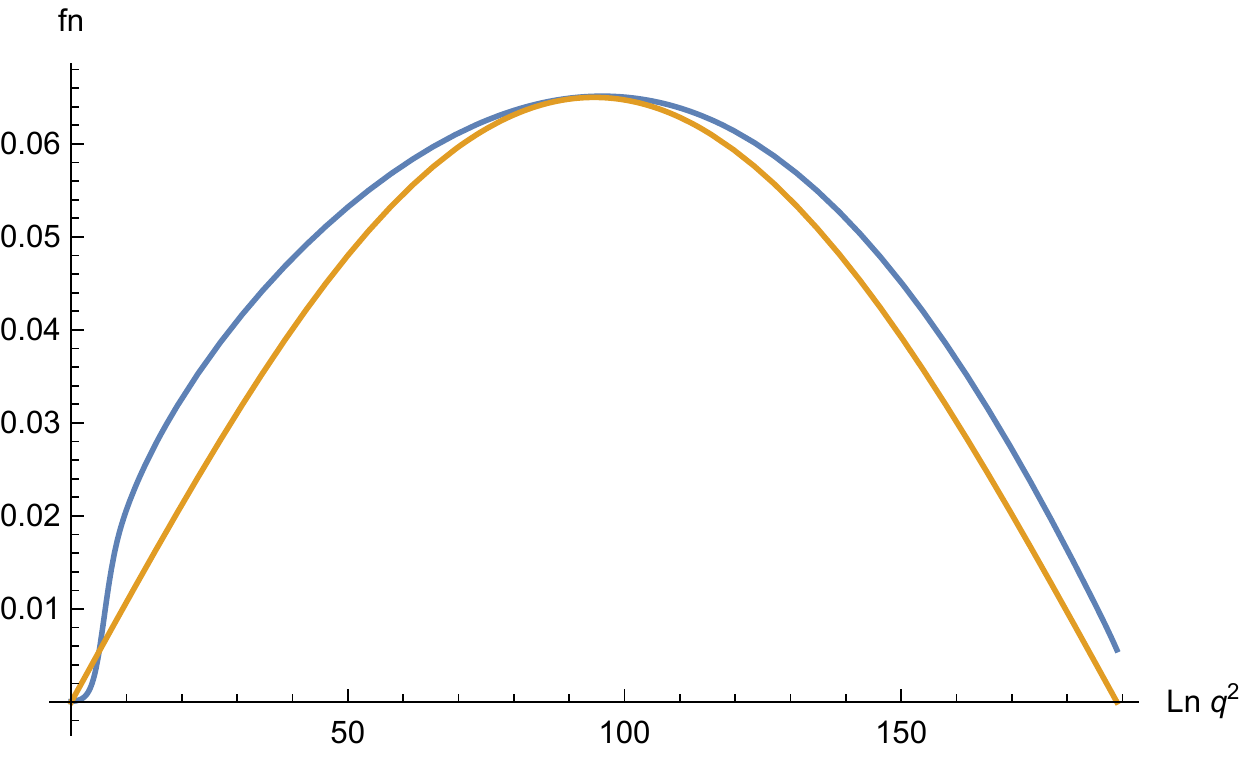}
	\end{minipage}
	\hspace{4cm}
	\begin{minipage}[t]{4cm} 
		\centering
		\includegraphics[scale=0.5]{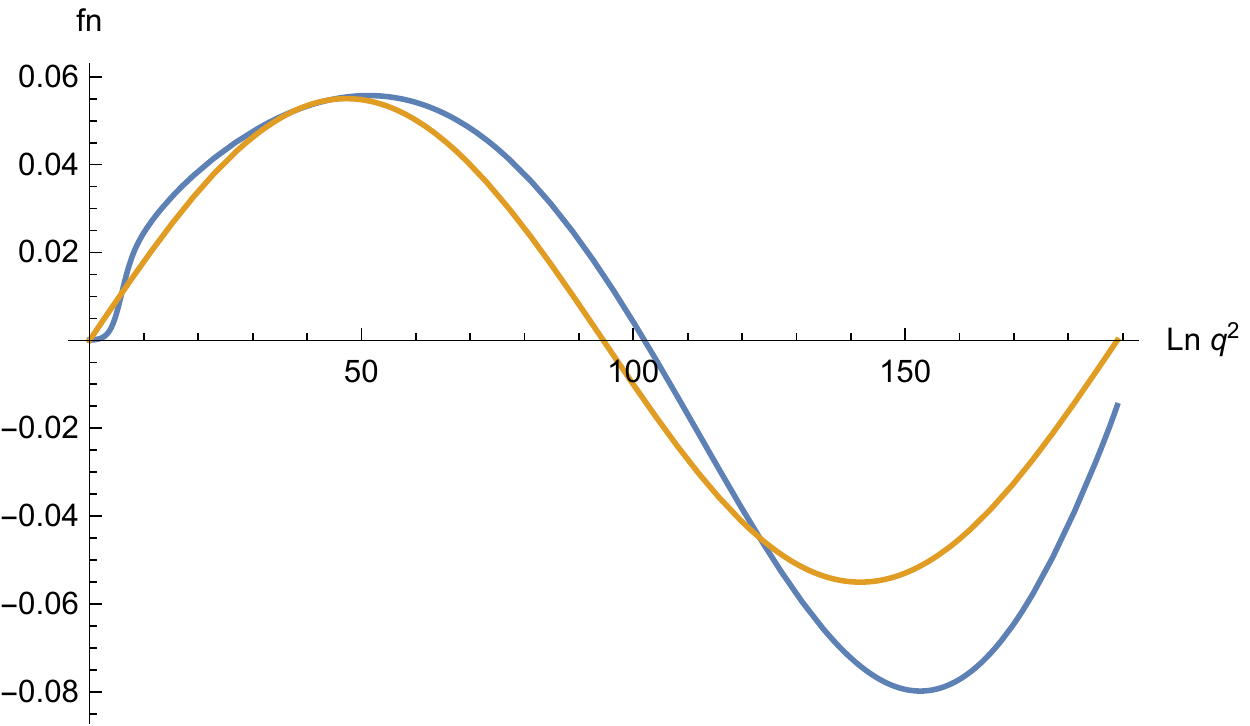}
	\end{minipage}
\hspace{3cm}
	\begin{minipage}[t]{4cm} 
		\centering
		\includegraphics[scale=0.5]{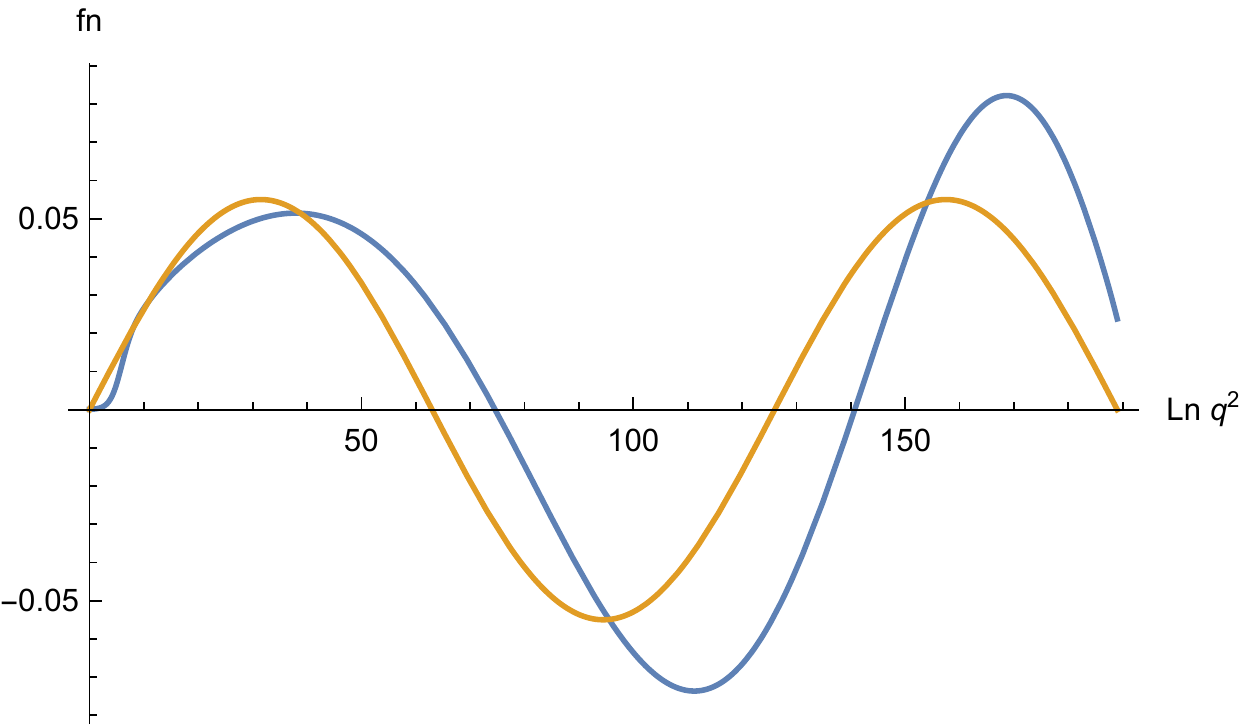}
	\end{minipage}
\caption{Comparation of the wave function $f_1$, $f_2$ and $f_4$ with    running  coupling constant and  mass regulator with the oscillatory  behavior of the  BFKL functions.}
\label{Fig7}
\end{figure}

\begin{figure}[H]
\begin{center}
\epsfig{file=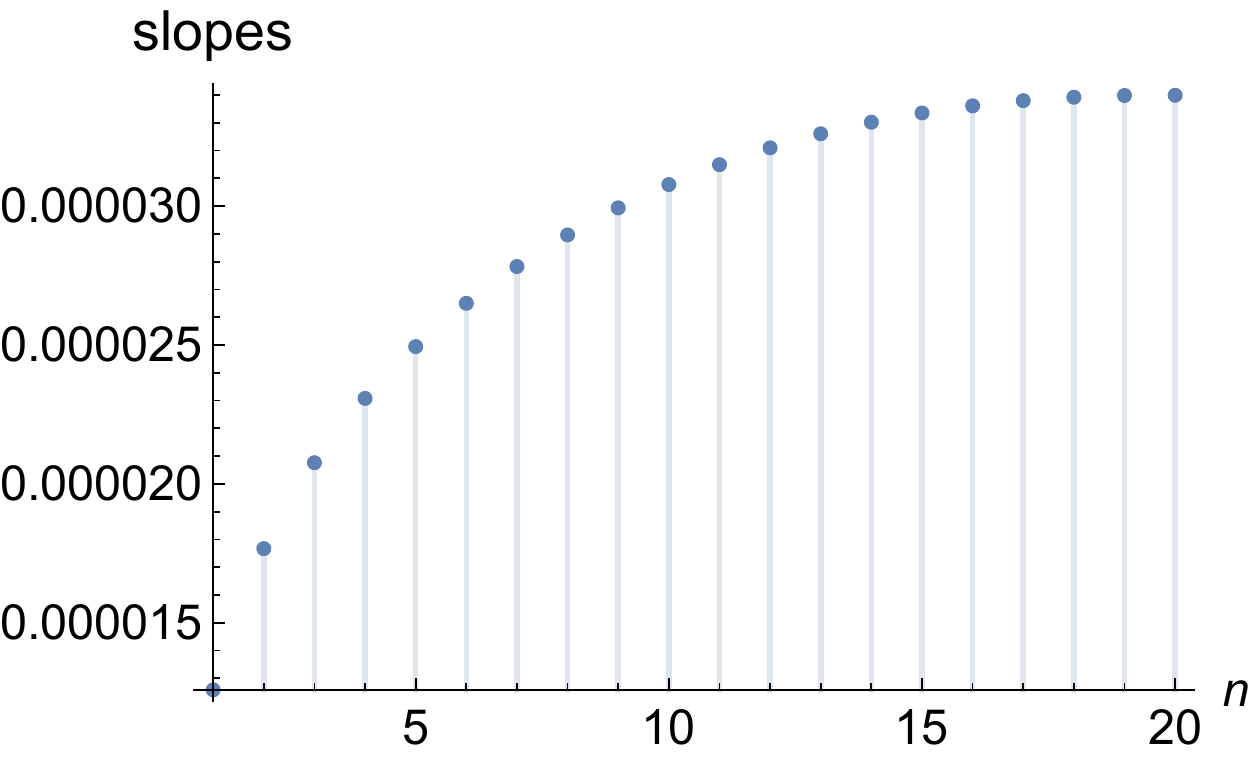,width=12cm,height=5cm}\\
\caption{$q^2$-slopes
\label{Fig8}} 
\end{center}
\end{figure}

Finally, in Fig.\ref{Fig8} we show the behavior of  the slopes from our numerical calculation.  One can observe that the slope increases  with $n$  but still remain smaller than a few times   $10^{-5}$. For the leading eigenvalues 
we find
\ba
E_1=&4 \times 10^{-6},&\alpha'_{1}= 1.26\times 10^{-5}\nonumber\\
E_2=&2.8\times 10^{-5},&  \alpha'_2=1.77 \times 10^{-5}  \nonumber\\
E_3=&7.4\times 10^{-5}, &\alpha'_3=2.08 \times 10^{-5} .
\ea
It is interesting to note that except for the leading state, numerically in the chosen unit the slopes are of the same order of the eigenvalues. 

\subsection{Dependence on the lattice size}

To further support our interpretation as a (fixed) cut in the energy plane, we note the following. In a continuum formulation of the BFKL eigenvalue equation, we expect the leading eigenvalue at exactly zero. For our finite lattice the leading eigenvalue turns out to be small and positive but nonzero, and for increasing lattice it should go to zero.
Indeed, for the much larger lattices with $ (t_{min},t_{max})=(-40,80)$, $(-40,100)$, $(-40,150)$, $(-40,170)$ (keeping $N_{step}=600$ fixed), one can see the decrease with increasing lattice size:
\ba 
&&E_1=3.9 \times 10^{-6},\,\,E_1=2.0 \times 10^{-6},\,\,E_1=6.0 \times 10^{-7},\,\,E_1=3.5 \times 10^{-7},\,\,\nonumber\\
&&E_2=2.7 \times 10^{-5},\,\,E_2=1.4 \times 10^{-5},\,\,E_2=4.4 \times 10^{-6},\,\,E_2=2.9 \times 10^{-6},\,\
\ea
More general, in Fig.\ref{Fig9} we show how all the eigenvalues decrease  as we increase the upper limit $ t_{max}$:
\begin{figure}[H]
\begin{center}
\epsfig{file=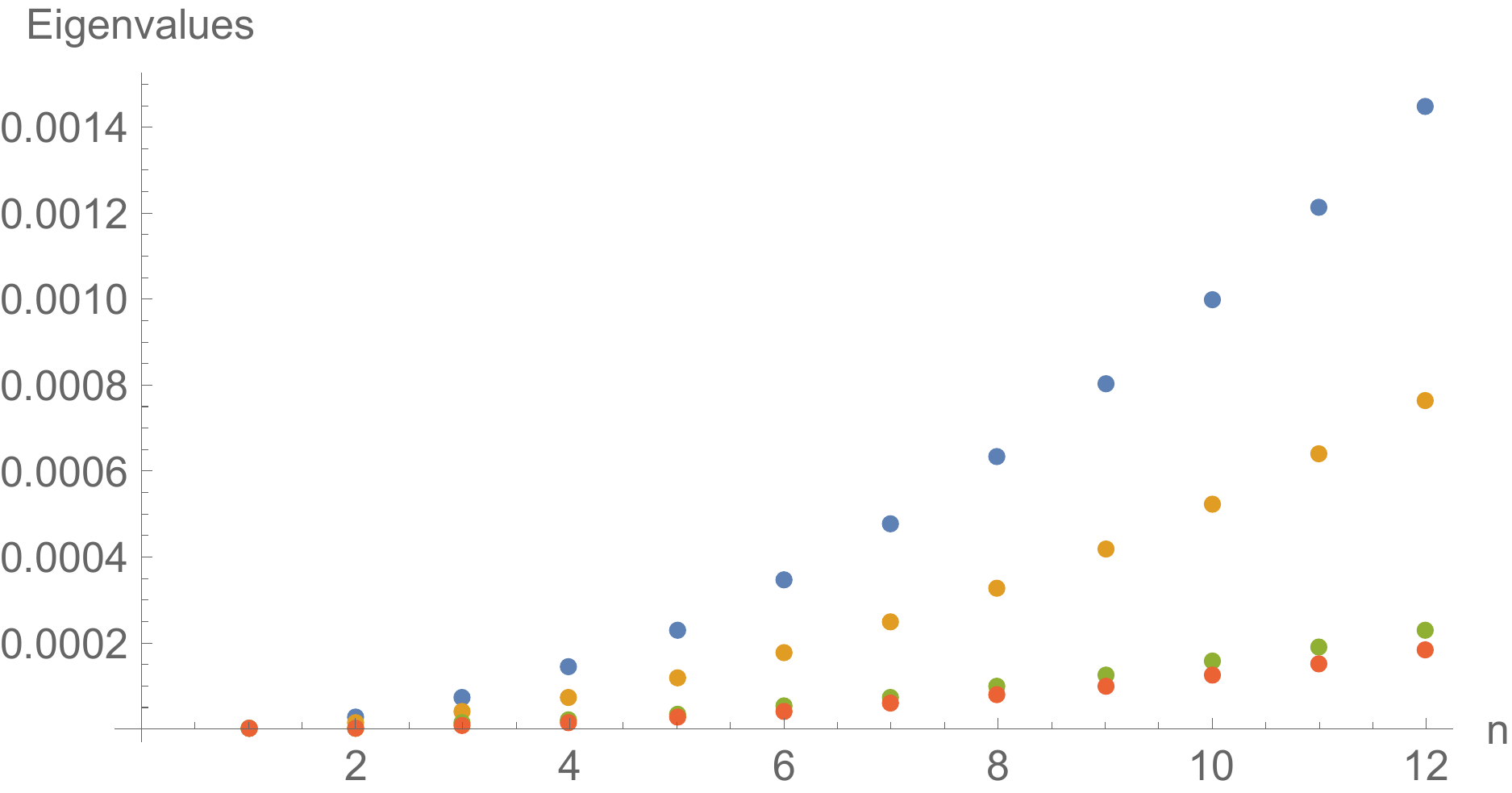,width=12cm,height=5cm}\\
\caption{Behavior of the eigenvalues when we increase the upper limit of the lattice: $t_{max}=80,100,150$ and $160$.
\label{Fig9}} 
\end{center}
\end{figure}
Simultaneous variation of the upper and lower limit lead to a further decrease of the eigenvalues, e.g. for  $N_{step}=600$ and $(t_{min},t_{max})=(-100,150 )$
\be
E_1=1.4 \times 10^{-7}, E_2=3.9 \times 10^{-6}.
\ee

Finally, for comparison we also vary $N_{step}=600, 800, 1000$ and $1200$,  keeping the lattice size constant $(t_{min}, t_{max})=(-40,80)$:
\ba
E_1=3.91 \times 10^{-6},\,\,E_1=3.92 \times 10^{-6},\,\,E_1=3.92 \times 10^{-6},\,\,E_1=3.92 \times 10^{-6},\,\,
\ea
This indicates that the numerical results are much less sensitive to $N_{step}$.

For the slope, we extend our  numerical analysis, keeping $N_{step}=600$ fixed. Comparing  $(t_{min},t_{max}) =(-40,80)$ and $(t_{min},t_{max}) =(-40,160)$
\ba
\alpha'_1&=&1.26 \times 10^{-5}\,\,, \alpha'_1= 2.3 \times 10^{-6}\nonumber\\
 \alpha'_2&=&1.77\times 10^{-5}\,\,, \alpha'_2=2.8 \times 10^{-6} 
\ea
we find analogous results also for the slope: they  decrease with increasing lattice size.

Similarly, for the wave functions (see Fig.\ref{Fig10}) with  $n$-nodes we find that with increasing lattice size the nodes 
move into the UV region. i.e. the location of the extrema  become larger with increasing lattice extension:
\begin{figure}[H]
\begin{center}
\epsfig{file=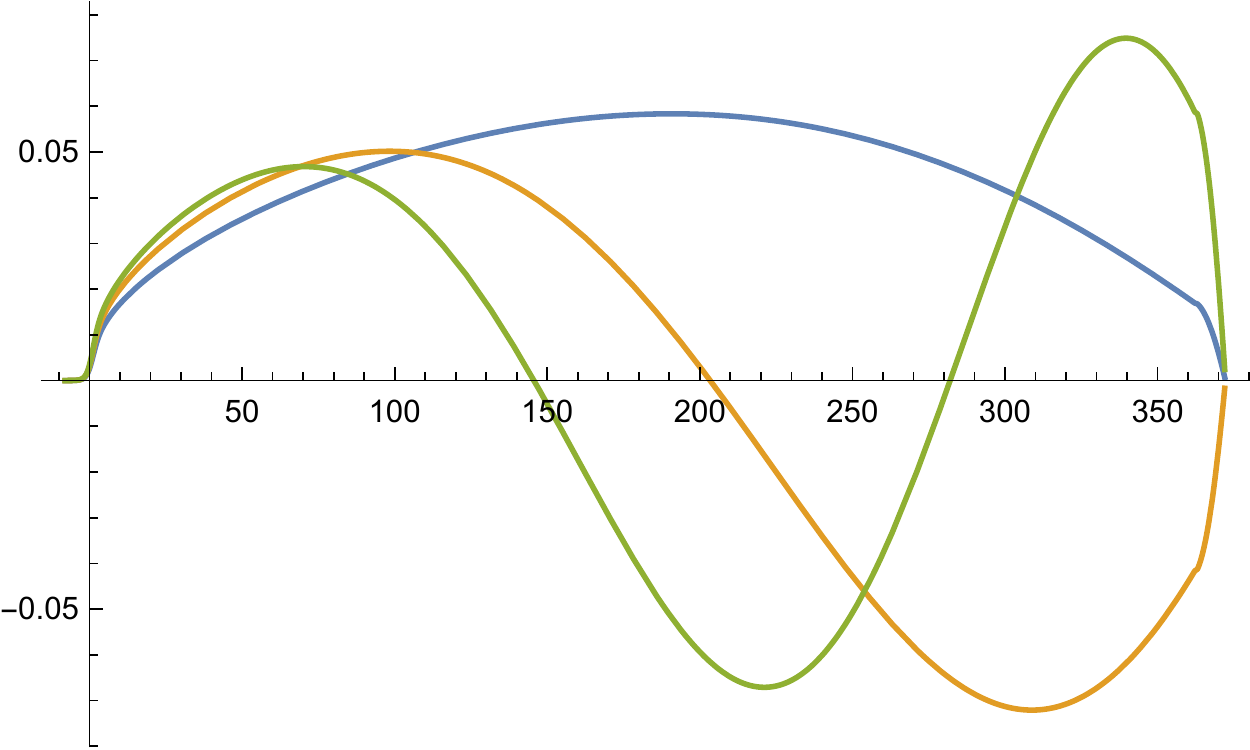,width=12cm,height=5cm}\\
\caption{Behavior of the wave function for a larger lattice limit $t_{max}=160$.
\label{Fig10}} 
\end{center}
\end{figure}

As to numerical values of the radii of the leading state, we again compare $(t_{min},t_{max}) =(-40,80)$ and $(t_{min},t_{max}) =(-40,160)$:
\ba
&<ln k^2> = 88, &<ln k^2> = 210.76 \nonumber\\
&r_1 = 6.8 \times 10^{18}\, GeV,\,\, & r_1 = 2.89 \times 10^{45} GeV .
\ea

All these results further support our conclusion that, at $q^2=0$,  our lattice formulation approximates the cut structure beginning at $E=0$ with wave functions extending to very large momenta or even to infinity. We see that lattice artifacts are under control.

\section{Summary and Outlook}
In this paper we have extended our previous analysis of the BFKL Pomeron to the Odderon case.
We have performed a numerical analysis of the BFKL equation for conformal spin=1, using a massive infrared regulator and the running coupling constant, introduced with a specific prescription. The main result of our work is that the spectrum remains essentially the same as it was without cutoff and with fixed coupling.
Let us note  that in a forthcoming publication \cite{MB and GPV2019},  M. Braun and  G. P. Vacca  have obtained very similar results: in this analysis a different infrared regulator is used which preserves the bootstrap condition of the BFKL equation. This supports the expectation that, in fact, the energy spectrum is fairly independent of the detailed form of the infrared regulator.  

It is important to stress the differences between the QCD Odderon and the Pomeron. As already stated in the 
introduction, the same procedure applied to the BFKL Pomeron equation leads to a discrete set of Pomeron states with intercepts above one and nonvanishing  $t$-slopes.  Moreover, the leading state is soft and its wavefunction has its support in the region of small transverse momenta. In contrast, the Odderon has no such discrete states for the leading (BLV) family of solutions: the fixed cut starts at $\omega=0$,  the wave functions have very small slopes, and their main support lies in the UV region. 
The most transparent way to study the effective momentum support seems to construct amplitudes integrating specific external particle impact factors (having characteristic scales) with the rapidity dependent Odderon Green's function.

It may be interesting to say a few words about the connection between the results of the present paper with the fixed point analysis performed in \cite{Bartels:2016ecw} in the soft  region. In this paper we 
have investigated the interaction of Pomeron and Odderon fields, assuming that, away from the infrared region, we have nonvanishing self-interactions of the Pomeron and interactions of Pomeron and Odderon, in particular a (real valued) Pomeron $\to$  2 Odderon vertex and an (imaginary) Odderon $\to$ Odderon+Pomeron vertex. We have found an infrared fixed point with two relevant (i.e. UV stable) directions. At this fixed point, both the Pomeron and the Odderon have intercept one and non vanishing slopes; the Odderon slope is slightly smaller than the Pomeron slope. When approaching this fixed point, in the parameter space of masses and interactions, from the IR stable 
directions  both intercepts initially are above one, and in the IR limit they then approach unity, the Odderon slightly faster than the Pomeron. If we associate the IR momentum cutoff $k$ with the radius $R$ of the scattering system $k^2 \sim 1/R^2$, and assume $R^2 = R_0^2 + 2 \alpha' \ln s$, we would expect that at large but finite energies 
the Odderon intercept would be slightly above unity, but smaller than the Pomeron intercept. 

When trying to connect these results with the findings of the present paper, one would be tempted to draw the following picture.
Starting in the UV region with the perturbative results for the Pomeron obtained in \cite{Bartels:2018pin} and for the Odderon described in the present  paper, one introduces interactions between Pomeron and Odderon fields and studies the RG flow as a function of the IR cutoff parameter $k$. In order to arrive at the IR fixed point described before,  these interactions have to lower the initial intercept above one of the BFKL Pomeron field, but also to modify the fixed-cut structure of the Odderon state. A study of this transition is in progress.\\ \\

\noindent
{\bf Acknowledgements:} J.B expresses his gratitude for support and hospitality of the Departamento de Fisica, Universidad Tecnica Federico Santa Maria, Valparaiso, Chile and for the support of the INFN and the hospitality of the Bologna University. C. C. thanks for the financial support from the grant FONDECYT 1191434 and 1180118, Chile.


\begin{thebibliography}{99}

\bibitem{Antchev:2017yns}
  G.~Antchev {\it et al.} [TOTEM Collaboration],
  arXiv:1812.04732 [hep-ex].

\bibitem{Antchev:2017dia}
  G.~Antchev {\it et al.} [TOTEM Collaboration],
  Eur.\ Phys.\ J.\ C {\bf 79} (2019) no.2,  103
  [arXiv:1712.06153 [hep-ex]].

\bibitem{Csorgo:2019fbf}
  T.~Csorg\"o [TOTEM Collaboration],
  EPJ Web Conf.\  {\bf 206} (2019) 06004
  [arXiv:1903.06992 [hep-ex]].

\bibitem{oddetotem1}  
E.~Martynov and B. Nicolescu
Phys. Lett.
{\bf B 778}, 414 (2018)


\bibitem{Nicolescu}  
L.~Lukazsuk and B.~ Nicolescu, Lett. Nuovo Cim. {\bf 8},  405 (1973) 

\bibitem{Donnachie:1983ff}
  A.~Donnachie and P.~V.~Landshoff,
  Phys.\ Lett.\  {\bf 123B} (1983) 345.

\bibitem{BFKL}
  L.~N.~Lipatov,
  Sov.\ J.\ Nucl.\ Phys.\  {\bf 23} (1976) 338
   [Yad.\ Fiz.\  {\bf 23} (1976) 642];\\
  
  E.~A.~Kuraev, L.~N.~Lipatov and V.~S.~Fadin,
  Sov.\ Phys.\ JETP {\bf 44} (1976) 443
   [Zh.\ Eksp.\ Teor.\ Fiz.\  {\bf 71} (1976) 840];\\
  
  E.~A.~Kuraev, L.~N.~Lipatov and V.~S.~Fadin,
  Sov.\ Phys.\ JETP {\bf 45} (1977) 199
   [Zh.\ Eksp.\ Teor.\ Fiz.\  {\bf 72} (1977) 377];\\
  
  I.~I.~Balitsky and L.~N.~Lipatov,
  Sov.\ J.\ Nucl.\ Phys.\  {\bf 28} (1978) 822
   [Yad.\ Fiz.\  {\bf 28} (1978) 1597].

\bibitem{Bartels:1980pe}
  J.~Bartels,
  Nucl.\ Phys.\ B {\bf 175} (1980) 365.

\bibitem{Kwiecinski:1980wb}
  J.~Kwiecinski and M.~Praszalowicz,
  Phys.\ Lett.\ B {\bf 94} (1980) 413.

  \bibitem{Janik:1998xj}
  R.~A.~Janik and J.~Wosiek,
  Phys.\ Rev.\ Lett.\  {\bf 82} (1999) 1092
  [hep-th/9802100].

 \bibitem{Bartels:1999yt}
  J.~Bartels, L.~N.~Lipatov and G.~P.~Vacca,
  Phys.\ Lett.\ B {\bf 477} (2000) 178
  [hep-ph/9912423].

\bibitem{Bartels:2001hw}
  J.~Bartels, M.~A.~Braun, D.~Colferai and G.~P.~Vacca,
  Eur.\ Phys.\ J.\ C {\bf 20} (2001) 323
  [hep-ph/0102221].
  
 \bibitem{ewert1}
 C.~Ewerz, The Odderon  in quantum chromodynamics,  arXiv:0306137 [hep-ph]

 \bibitem{Braun:1996tc}
  M.~Braun, G.~P.~Vacca and G.~Venturi,
  Phys.\ Lett.\ B {\bf 388} (1996) 823
  [hep-ph/9605304].

\bibitem{Kowalski:2017umu}
  H.~Kowalski, L.~N.~Lipatov, D.~A.~Ross and O.~Schulz,
  Eur.\ Phys.\ J.\ C {\bf 77} (2017) no.11,  777
  [arXiv:1707.01460 [hep-ph]].

\bibitem{Kowalski:2015paa}
  H.~Kowalski, L.~N.~Lipatov and D.~A.~Ross,
  Eur.\ Phys.\ J.\ C {\bf 76} (2016) no.1,  23
  [arXiv:1508.05744 [hep-ph]].

\bibitem{Kowalski:2014iqa}
  H.~Kowalski, L.~Lipatov and D.~Ross,
  Eur.\ Phys.\ J.\ C {\bf 74} (2014) no.6,  2919
  [arXiv:1401.6298 [hep-ph]].

\bibitem{Levin:2014bwa}
  E.~Levin, L.~Lipatov and M.~Siddikov,
  Phys.\ Rev.\ D {\bf 89} (2014) no.7,  074002
  [arXiv:1401.4671 [hep-ph]].
  
  \bibitem{Levin:2015noa}
  E.~Levin, L.~Lipatov and M.~Siddikov,
  Eur.\ Phys.\ J.\ C {\bf 75} (2015) no.11,  558
  [arXiv:1508.04118 [hep-ph]].
  
\bibitem{Levin:2016enb}
  E.~Levin, L.~Lipatov and M.~Siddikov,
  Phys.\ Rev.\ D {\bf 94} (2016) no.9,  096004
  [arXiv:1608.03816 [hep-ph]].

\bibitem{Bartels:2018pin}
  J.~Bartels, C.~Contreras and G.~P.~Vacca,
  JHEP {\bf 1901} (2019) 004
  [arXiv:1808.07517 [hep-ph]].
  
\bibitem{Bartels:2015gou}
  J.~Bartels, C.~Contreras and G.~P.~Vacca,
  JHEP {\bf 1603} (2016) 201
  [arXiv:1512.07182 [hep-th]].
  
\bibitem{Bartels:2016ecw}
  J.~Bartels, C.~Contreras and G.~P.~Vacca,
  Phys.\ Rev.\ D {\bf 95} (2017) no.1,  014013
  [arXiv:1608.08836 [hep-th]].
  
\bibitem{Bartels:2012sw}
  J.~Bartels, V.~S.~Fadin, L.~N.~Lipatov and G.~P.~Vacca,
  Nucl.\ Phys.\ B {\bf 867} (2013) 827
  [arXiv:1210.0797 [hep-ph]].

\bibitem{Bartels:2013yga}
  J.~Bartels and G.~P.~Vacca,
  Eur.\ Phys.\ J.\ C {\bf 73} (2013) 2602
  [arXiv:1307.3985 [hep-th]].

 \bibitem{MB and GPV2019}
 M.~Braun  and G.~P.~Vacca, to be publish and private communication.

  
\end{thebibliography}
\end{document}